\def\X{{\rm X}^2}
    \def\Y{{\rm Y}^2}
\def\xxz{{\em xxz\/}}
\documentclass[10pt,prb,aps,twocolumn]{revtex4} 
\usepackage{graphicx}
\usepackage{amssymb}

\advance\textfloatsep by -0.25in
\advance\abovecaptionskip by -0.1in
\advance\floatsep by -0.2in
\advance\dblfloatsep by -0.5in

\advance\textheight by 0.03in
\begin{document}

\title{Scalable design of tailored soft pulses for coherent control}

\author{Pinaki Sengupta}
\affiliation{Department of Physics, University of California, Riverside, CA
  92521}
\author{Leonid P. Pryadko} 
\affiliation{Department of Physics, University of California, Riverside, CA
  92521}

\date{February 3, 2005}

\begin{abstract}
We present a scalable scheme to design optimized soft pulses and
pulse sequences for coherent control of interacting quantum
many-body systems.  The scheme is based on the cluster expansion and
the time dependent perturbation theory implemented numerically.
This approach offers a dramatic advantage in numerical efficiency,
and it is also more convenient than the commonly used Magnus
expansion, especially when dealing with higher order terms.  We
illustrate the scheme by designing $2$nd-order self-refocusing $\pi$-pulses and
a $6$th-order $8$-pulse refocusing sequence for a chain of qubits with
nearest-neighbor couplings.  We also discuss the performance of
soft-pulse refocusing sequences in suppressing decoherence due to
low-frequency environment.
\end{abstract}


\maketitle

A control of coherent evolution of quantum systems is increasingly
important in a number of research fields and applications.  Such
control has long been a staple in nuclear magnetic resonance (NMR)
spectroscopy, where determination of the structure of complex
molecules like proteins require the application of long sequences of
precisely designed radio-frequency (r.f.)
pulses\cite{slichter-book}. 
Recently, coherent 
control (CC) has emerged as an important part of quantum information
processing (QIP), spurring numerous studies on general properties and
specific design of pulses and pulse sequences for
application in NMR-based\cite{chuang-2001} 
and other potential
implementations\cite{platzman-dykman-seddighrad} 
of
quantum computers (QCs).  This progress is closely followed by
applications of coherent control in atomic physics\cite{weiman-1999},
quantum optics\cite{osborne-coontz-2002}, laser induced chemical
reactions\cite{goswami-review}, data
communications\cite{laser-data-comm}, and biomedical
applications\cite{bio-laser}.

The precision required for QIP in particular, and for contemporary
applications of CC in general, is achieved most readily using shaped
(also, soft), typically narrow-band, pulses.  When properly
constructed, such pulses allow excitation to be limited to a
particular set of modes which results in better control fidelity and
reduced incoherent losses (e.g., heating).  The latter is especially
important for putative solid-state QC implementations which are
proposed to operate at cryogenic temperatures.  Additionally, as we
also discuss in this work, refocusing with carefully designed
high-order pulses and pulse sequences can offer significantly better
protection against non-resonant decoherence sources (e.g.,
low-frequency phonons) as compared to lower-order sequences.

Over some forty years the shaped pulses were utilized in NMR, a number
of analytical and numerical schemes were suggested for their
design\cite{slichter-book}. 
Most (although not all, see Ref.~\onlinecite{abramovich-vega-1993})
rely on the average Hamiltonian theory, a perturbative scheme based on
the cumulant (Magnus) expansion for the evolution operator.  The
expansion is done around the evolution in the applied controlling
fields, while the chemical
shifts\cite{warren-herm} 
(resonant-frequency offsets) and, ideally, inter-spin couplings are
treated perturbatively.  The main drawback of the Magnus expansion [see
Eq.~(\ref{eq:cumulant}) below] for numerics is the multiple
integration appearing in higher orders; its use in actual
calculations was almost always limited to quadratic order.

The alternative scheme found in the literature is a simulation
involving the full Hamiltonian of a quantum
system\cite{sanders-1999-complex-instruct}. 
In particular, such a calculation was done 
to optimize pulse shapes for a three-level Hamiltonian of a
current-biased superconducting qubit\cite{steffen-2003}.  This
approach guarantees precision of custom-designed  shapes, but it
obviously lacks scalability, as the computational difficulty grows
exponentially with the size of the system. 

In this paper, we present an efficient scalable scheme to design
high-order soft 
pulses and soft-pulse sequences for controlling quantum many body
systems.  Instead of using the Magnus expansion or other form of the
effective Hamiltonian theory, we rely directly on the time-dependent
perturbation theory implemented numerically.  This allows an easy 
extension to higher orders (up to $9$\,th in this work), at the same
time preserving the benefits of the cluster
theorem\cite{Domb-Green-book} which limits the size of the
system to be analyzed.  The high order calculation allows a straightforward
classification of pulse sequences by order $K$, the number of terms in
the time dependent perturbation series for which the control remains
perfect.

To be specific, and also as an illustration, we consider a
quantum spin chain with each spin (qubit)
individually controlled and  ``always-on''
nearest-neighbor (n.n.)\ interactions.  Assuming the Ising coupling is
dominant, $J_z\gtrsim J_\perp$, we construct a family of
one-dimensional $\pi$-pulses\cite{endnote} 
with different degrees of 
self-refocusing with respect to the $J_z$ coupling.  The duration of a
pulse, $\tau$, is chosen to be fixed, so as to allow parallel execution of
quantum gates in different parts of the system.  To reduce the
spectral width of an arbitrary sequence of such pulses we also require
a number of derivatives of the controlling fields to vanish at the
ends of the cycle.

We show that thus designed pulses work as drop-in replacement of hard
pulses, and compare their performance in several pulse sequences and
composite pulses with that of two commonly used  shapes.  In particular,
we present an eight-pulse refocusing sequence of order $K=6$ for the
quantum Ising model [refocusing errors scale as $(J_z\tau)^6$], order
$K=2$ for general \xxz\ model, where, in addition, each spin has an
order $K\ge2$ protection against phase decoherence due to low-frequency
environment (TAB.~\ref{tab:pulses}).

\begin{table}[thbp]
  \centering
  \begin{tabular}[c]{|c|llll|llll|llll|}
    Model &\multicolumn{3}{r}{Ising}& &\multicolumn{3}{r}{\xxz}& &
    \multicolumn{3}{r}{bath}&\\ \hline 
    Sequence                     
    & {\bf 1} & {\bf 2} & {\bf 4} & {\bf 8} 
    & {\bf 1} & {\bf 2} & {\bf 4} & {\bf 8} 
    & {\bf 1} & {\bf 2} & {\bf 4} & {\bf 8}\\\hline 
    Gauss [\onlinecite{warren-herm}b] 
    & 0 & 1 & 1 & 2 & 0 & 0 & 0 & 1 & 0 & 1*& 0 & 1 \\
    Herm [\onlinecite{warren-herm}a], S$_1$
    & 1 & 1 & 3 & 4 & 0 & 0 & 1 & 1 & 1*& 1*& 1 & 1\\
    Q$_1$                        
    & 2 & 3 & 5 & 6 & 0 & 0 & 1 & 2 & 2*& 2*& 2 & 3\\
  \end{tabular}
  \caption{Order $K$ determining the scaling ($\propto \tau^{K}$) of the gate
     errors  with the pulse duration $\tau$ for different refocusing
    sequences.  
    {{\bf 1}}:\ a single $\pi$-pulse along the $x$-axis on odd sites, ``$
    \X_1$''; {{\bf 2}}:\ two $\pi$-pulses along the $x$-axis, odd
    sites only, $\X_1\X_1$; {{\bf 4}}:\ %
    $\X_1\Y_2\overline\X_1\overline\Y_2$ (bars for negative pulses,
    subscripts denote odd or  
    even sites); {{\bf 8}}:\ %
    $\X_1\Y_2\overline\X_1\overline\Y_2\overline\Y_2\overline\X_1\Y_2\X_1$.    
    Asterisks mark odd-site refocusing.      See text for description
    of the three models.}
  \label{tab:pulses}
\end{table}

We consider the following simplified Hamiltonian
\begin{equation}
  \label{eq:ham-total}
   H(t)= H_{\rm C}(t)+ H_{\rm S}+ H_{V}(t)+ H_{\sigma},
\end{equation}
with the first (main) term  due to individual control fields, 
\begin{equation}
  \label{eq:ham-control}
   H_{\rm C}(t)={1\over2}{\textstyle\sum}_n \bigl[ V^x_n(t)\,
  \sigma^x_n+V^y_n(t)\, \sigma^y_n\bigr] , 
\end{equation}
where $\sigma_n^\mu$, $\mu=x,y,z$, are the usual Pauli matrices
for the $n$-th qubit (spin) of the 1D chain.  
The other terms describe the
interactions between the qubits (n.n.\ \xxz),
\begin{equation}
  \label{eq:ham-internal}
   H_{\rm S}={1\over4}\sum_{\langle n,n'\rangle} \bigl[
  J_{n,n'}^z \sigma^z_n
  \sigma^z_{n'}+ 
  J_{n,n'}^\perp (\sigma^x_n \sigma^x_{n'}+\sigma^y_n
  \sigma^y _{n'})\bigr],
\end{equation}
and the coupling with the oscillator thermal bath,
\begin{equation}
  \label{eq:ham-env}
  H_{V}(t)=   {\textstyle\sum}_{n\mu}
  A_n^\mu\,V_n^\mu(t),\quad
  H_\sigma= {1\over2}{\textstyle\sum}_n  B_{n}^{\mu} \,  \sigma^\mu_n.
\end{equation}
In Eq.~(\ref{eq:ham-env}), $A_n^\mu\equiv A_n^\mu(p_i,q_i)$
account for the possibility of a direct coupling of the controlling
fields $V_n^\mu$ with the bath variables $q_i$, $p_i$, while
$B_n^\mu\equiv B_n^\mu(p_i,q_i)$ describe the usual coupling of the
spins with the oscillator bath.  Already in the linear response
approximation, the bath couplings~(\ref{eq:ham-env}) produce a
frequency-dependent renormalization of the control Hamiltonian $H_{\rm
  C}(t)$ [Eq.~(\ref{eq:ham-control})], as well as the thermal bath
heating via the dissipative part of the corresponding response
function.  Both effects become more of a problem with increased
spectral width of the controlling signals $V_n^\mu$.  In this work we
do not specify the explicit form of the coupling $H_V(t)$.  Instead,
we minimize the spectral width of the constructed pulses.

{\bf Closed system}:\ In a 
qubit-only system with the Ha\-mil\-tonian 
$H(t)=H_{\rm C}(t)+ H_{\rm S}$, the effect of the applied fields is
fully described by the evolution operator $U(t)$, 
\begin{equation}
  \label{eq:evol-oper}
  \dot U(t) =-i\, [ H_{\rm C}(t)+ H_{\rm S}]\, U(t),\quad U(0)=\mathbb{I}.
\end{equation}
As usual, the time-dependent perturbation theory is introduced by
separating out the bare evolution operator, 
\begin{equation}
  \label{eq:evol-bare}
  U(t)=U_0(t)\,R(t),\quad
  \dot U_0(t) =-i H_{\rm C}(t)\, U_0(t) .
\end{equation}
Then, the operator $R(t)$ obeys the equation
\begin{equation}
  \dot R(t)=-i \tilde H_S(t) R(t), \quad 
  \tilde H_{\rm S}(t)\equiv U_0^\dagger (t)\,   H_{\rm S}\, U_0(t),
  \label{eq:evol-oper-inter}
\end{equation}
which can be iterated to construct the standard  expansion $R(t)
  =\mathbb{I}+R_1(t) +R_2(t) +\ldots$ in powers of $(t\,H_{\rm
  S})$, 
\begin{equation}
  \label{eq:pert-expansion}
    \dot R_{n+1}(t)=-i \tilde H_S(t) R_{n}(t), \quad
    R_{0}(t)={\mathbb I}.
\end{equation}
For a finite system of $n$ qubits and a given maximum order $K$ of the
expansion, Eqs.~(\ref{eq:evol-bare}), (\ref{eq:evol-oper-inter}), and
(\ref{eq:pert-expansion}) are a set of coupled first order ordinary
differential equations for the $2^n\times 2^n$ matrices $U_0$, $R_1$,
$R_2$, \ldots, $R_K$, and can be integrated efficiently using any of
the available extrapolation schemes.  Obviously, for a given system,
solving the full equations~(\ref{eq:evol-oper}) is simpler by a factor
of at least $(K+1)$.  However, it is the analysis of the perturbative
expansion that is the key for achieving the scalability of the
results.

The standard Magnus expansion can be readily obtained by integrating
Eqs.~(\ref{eq:pert-expansion}) formally and rewriting the result in
terms of cumulants,
\begin{eqnarray}
  \label{eq:cumulant}
  \nonumber
  R(t)&=&\exp( C_1+C_2+\cdots), \;\,C_1=-i\int_0^t dt_1 \tilde H_{\rm
  S}(t_1),\\ 
  C_2&=&-{1\over 2}\int_0^t dt_2\int_0^{t_2} dt_1 [\tilde H_{\rm
  S}(t_1),\tilde H_{\rm S}(t_2)],\;\,\cdots 
\end{eqnarray}
Generally, the term $C_k$ contains a $k$-fold integration of the
commutators of the rotating-frame Hamiltonian $\tilde H_{\rm S}(t_i)$
at different time moments $t_i$ and has an order $(t H_{\rm S})^k$.
The advantage of the cumulant expansion is that it does not contain
the disconnected terms arising from different parts of the system.
For an arbitrary lattice model of the form (\ref{eq:ham-internal}),
with bonds representing the qubit interactions, the terms contributing to
$k$-th order can be represented graphically as connected
clusters involving up to $k$ lattice
bonds; generally such clusters 
cannot have more than $n=k+1$ vertices.  Thus, to obtain the exact
form of the expansion up to and including $K$-th order, one needs to
analyze all distinct clusters with up to $K+1$
vertices.  For an 
infinite chain with n.n.\ couplings, these are finite chains with up
to $K$ bonds and $K+1$ vertices.

The discussed cluster theorem\cite{Domb-Green-book} appears
to offer a distinct advantage to 
the Magnus expansion compared with the regular perturbation theory.
On the other hand, evaluation of multiple integrals
(\ref{eq:cumulant}) directly is
computationally challenging, which limits the use of higher-order
Magnus expansions for numerics.  We note, however, that the order-$K$
universal self-refocusing condition 
$C_1=\ldots=C_K=0$ 
is formally
equivalent to
\begin{equation}
R_1=R_2=\ldots=R_K=0.
\label{eq:self-refocusing}
\end{equation}
The matrices $R_k$ in the latter condition are much easier to
evaluate numerically using
Eqs.~(\ref{eq:evol-bare})---(\ref{eq:pert-expansion}).  Yet, the
benefits of the cluster theorem remain: to $K$-th order only clusters
with up to $K+1$ vertices need to be analyzed.

We implemented the described scheme using the standard fourth-order
Runge-Kutta algorithm for solving coupled differential equations, and
the GSL library\cite{gsl-ref} for matrix operations.  The coefficient
optimization was done using a combination of simulated annealing and
the steepest descent method.  The trial pulse shapes were encoded in
terms of their Fourrier coefficients,
\begin{equation}
  V(t+\tau/2) =A_0+\sum_m A_m \cos (m\Omega t)+B_m\sin(m\Omega t),
  \label{eq:pulse-fourrier}
\end{equation}
where the angular frequency $\Omega=2\pi/\tau$ is related to the full
pulse duration $\tau$.  The target function for single-pulse
optimization included the sum of the magnitudes squared of the matrix
elements of the zeroth-order mismatch matrix $[U_0(\tau)-U_{\rm
  target}]$, and of the matrices $R_k(\tau)$, $k=1,\ldots,K$.  The
minimization continued until these contributions went down to zero
with the numerical precision (typically, eight digits or more).

As the simplest application of the formalism, we designed a number of
inversion ($\pi$-) pulse shapes\cite{endnote}, self-refocusing to various degrees
with respect to the Ising interaction; their coefficients are listed
in TAB.~\ref{tab:coeff}.  To reduce the bandwidth of an arbitrary
sequence of such pulses, we required additionally that the function
(\ref{eq:pulse-fourrier}) vanishes along with a number of its
derivatives $V^{(l)}(t)$, $l=1, 2, \ldots, 2L-1$, at the ends of the
interval, $t=0,\tau$.

\begin{table*}[htbp]
  \begin{tabular}[c]{c|c|c|c|c|c|c}
    & $A_0$& $A_1$& $A_2$& $A_3$& $A_4$ & $A_5$\\
    \hline
    $S_1$& 0.5&  -1.2053194466& 0.4796460175 &   0.2256734291 & \\
    $S_2$& 0.5 & -1.1950755990& 0.7841246569 &  0.0738054432 &  -0.1628545011\\
    $Q_1$& 0.5 &  -1.1374003264 &  1.5774784244&  -0.6825954606 &
    -0.2574826374\\
    $Q_2$& 0.5 & -1.0965122417 &  1.5309957409 &  -1.1470791601 &
    0.0020722004 &  0.2105234605 
  \end{tabular}\centering
  \caption{Fourrier coefficients for the constructed pulses, see
    Eq.~(\ref{eq:pulse-fourrier}).  Shapes $S_L$ and $Q_L$
    respectively are first
    ($K=1$) and  second ($K=2$) order self-refocusing
    inversion pulses for the Ising spin coupling; the fixed-time 
    errors scale with the duration of the pulse as $\propto (\tau
    J_z)^K$ [cf.\ Eq.~(\ref{eq:ham-internal})].  These shapes
    have $2L$  derivatives vanishing at the ends of the interval.}
  \label{tab:coeff}
\end{table*}

These shapes can work in known high-order pulse
sequences\cite{brown-harrow-chuang-2004} 
as a
drop-in replacement of hard (or short Gaussian) pulses.  We note that
in our setup there is no gap between subsequent pulses, the pulses
follow back to back with the repetition period $\tau$.  The system is
``focused'' at the end of each time interval.  Such a scheme with a
common ``clock'' time $\tau$ is convenient, e.g., for parallel
execution of quantum gates in different parts of the system.  For each
qubit, various pulses (or intervals of no signal) can be executed in
sequence.  The performance of such sequences can be analyzed in the
same manner as that of a single pulse.  Namely, we integrate
Eqs.~(\ref{eq:evol-bare}), (\ref{eq:evol-oper-inter}), and
(\ref{eq:pert-expansion}) over the full duration $t$ of the pulse
sequence; the order of the sequence is the number $K$ of the exactly
cancelled terms in the perturbative expansion of 
$R(t)$.
After $N=t/\tau$ steps, the error in the unitary evolution matrix
would scale as $\propto N \tau^{K+1}= t \tau^K$; the corresponding
gate fidelity (defined as the probability of error, either average or
maximum) would scale as $1-\mathcal{O}(\tau^{2K})$.

In Tab.~\ref{tab:pulses}, we illustrate the quality of the obtained
pulses by comparing their performance in several refocusing sequences
for different models. ``Ising'': the Ising-only interaction
[Eq.~(\ref{eq:ham-internal}) with all $J^\perp=0$]; ``\xxz'': the
\xxz\ spin chain with both $J^z$ and $J^\perp$ non-zero; ``bath'':
Ising spin chain coupled to a thermal bath generating slow (compared
to $\tau$) phase modulation, simulated as $H_\sigma$
[Eq.~(\ref{eq:ham-env})] with random time-independent coefficients
$B_n^z$ (see further discussion on open systems below).  The pulse
sequences are listed in the caption; these are ``best'' sequences at
given length for all pulse shapes found by exhaustive search
(high-order sequences\cite{brown-harrow-chuang-2004} equivalent for
hard pulses do not necessarily have equal orders here).  The fact that
such a brute-force optimization approach works is entirely due to the
efficiency of the method.

The most interesting is the length-8 sequence
``$\X_1\Y_2\overline\X_1\overline\Y_2\overline\Y_2\overline\X_1\Y_2\X_1$'',
where $\X_1$ is a $\pi$-pulse in $x$-direction applied on every odd
site, $\overline\Y_2$ is a $\pi$-pulse in negative $y$-direction on
even sites, etc.  This sequence is the best among the length-eight
sequences for both the Ising and the \xxz\ ($J^\perp\neq0$) models,
and, additionally, it protects every qubit from phase decoherence due
to low frequency noise.  Our second-order self-refocusing pulses are
clearly advantageous, especially if the Ising coupling is
dominant.   The corresponding errors scale as
$(J_z\tau)^6$ compared with that for the standard (first-order)
Hermitian pulse where gate error scales as $(J_z\tau)^4$ [the gate
fidelities differ from unity by
$\mathcal{O}\biglb((J_z\tau)^{12}\bigrb)$ and
$\mathcal{O}\biglb((J_z\tau)^{8}\bigrb)$ respectively].

These pulses were designed for use in systems with dominant Ising
coupling, and this is the situation where they are most useful as a
replacement of, say, Gaussian pulses.  For example, when the pulse
$Q_1$ along with analogously designed second order $\pi/2$ and $2\pi$
pulses were used to simulate the BB$_1$ composite
pulse\cite{wimperis-1994} designed to compensate for amplitude errors
to third order, the results for a single spin were essentially
identical to those with Gaussian pulses, with errors cubic in the
amplitude mismatch.  However, when used in an Ising chain, the
performance of the BB$_1$ sequence with Gaussian pulses
deteriorated linearly in $J_z\tau$ already with zero
amplitude mismatch, while for our second-order pulses the additional
error was smaller, scaling as  the product of $(J_z\tau)$ and the amplitude
mismatch.  Clearly, if the two sources of errors are comparable,
combining high-accuracy BB$_1$ composite pulse and the second order pulses
may be superficial; simpler pulse sequence and/or pulses with first
order compensation could give a comparable accuracy.  

{\bf Open systems}:\ Qualitatively, the effect of the refocusing
pulses on the thermal bath coupling $H_\sigma$ [Eq.~\ref{eq:ham-env}]
can be most readily understood in the rotating frame defined by the
bare evolution operator $U_0(t)$ [Eq.~\ref{eq:evol-bare}],
\begin{equation}
  \tilde H_\sigma(t)=U_0^\dagger(t) H_\sigma U_0={1\over2}{\textstyle\sum_n}
  B_n^\mu(p_i,q_i)
  Q_n^{\mu\mu'}(t)\sigma_n^{\mu'}.\label{eq:rotating-hsigma}
\end{equation}
For refocusing, the rotation matrices $Q_n^{\mu\mu'}(t)$ are periodic
with the full sequence period $\tilde\tau$; they can be written as a
sum of harmonics with the main frequency $\tilde\Omega=2\pi /\tilde \tau$,
\begin{equation}
  Q_n^{\mu\mu'}(t)=\sum_m C_{nm}^{\mu\mu'}e^{-i\tilde\Omega_m
    t}, \quad \tilde\Omega_m\equiv m\tilde\Omega.
  \label{eq:rotation-matrix-harmonics}
\end{equation}
The constant-field first-order refocusing condition (average
Hamiltonian vanishes to leading order) is equivalent to a
cancellation of some linear combinations of $C_{n0}^{\mu\mu'}$
(e.g., $C_{n0}^{z\mu'}=0$ 
for phase noise assumed for thermal bath in Tab.~\ref{tab:pulses}).
As a result, the environmental modes at low frequencies get modulated
and are effectively replaced by those at higher frequencies,
$\omega\to\omega+\tilde\Omega_m$, $m\neq0$, leading to a significant
reduction of the decoherence caused by resonant decay
processes\cite{dykman-1979,kofman-kurizki-2001,shiokawa-lidar-2004}.
On the other hand, fast modes are mostly unaffected; modulation has
essentially no effect on a ``fast'' (e.g., $\delta$-correlated)
thermal bath.

Quantitatively, the effect of refocusing can be understood with the
quantum kinetic equation (QKE) in the rotating frame, with the kernel
accurate at least to order $K$ to analyze order-$K$
refocusing\cite{pryadko-qke}.  For large $\tilde\Omega$, the
density-matrix dynamics separates onto sectors with frequencies around
$\tilde\Omega_m$.  The slow sector, $m=0$, carries the main part of
the total weight, with that of the remaining (generally,
rapidly-decaying) sectors totalling $\sim\Delta(0)/\tilde\Omega^2$,
where $\Delta(t-t')\equiv \bigl\|\langle
B^\mu(t)B^{\mu'}(t')\rangle\bigr\|$ is a norm of the correlation
matrix of the fluctuating field.  Only the dynamics in the slow sector
is protected by the refocusing.  In particular, the analysis of the
QKE with the leading second-order kernel shows that already with
first-order ($K=1$) constant-field refocusing direct decay processes
require excitations at frequencies $\omega\gtrsim\tilde\Omega$, which
may dramatically reduce the dissipative part of the QKE kernel.  The
non-resonant reactive processes are also suppressed: the rate of phase
errors is $\sim \Delta(0)/\tilde\Omega$ with $K=1$ refocusing and
$\sim |\Delta''(0)|/\tilde\Omega^3$ (primes denote time derivatives)
with $K=2$ refocusing, as, e.g., for length-8 sequence in
Tab.~\ref{tab:pulses}.  Generally, these results\cite{pryadko-qke}
apply equally for soft- and 
hard-pulse refocusing, and are consistent with established results on
kinetics of  few-level
systems in r.f.\ field\cite{dykman-1979},  and with the properties of 
hard pulse sequences for low-frequency
environment\cite{shiokawa-lidar-2004,facchi-nakazato-2004}. 

To conclude, we presented an efficient scheme for designing high order
soft pulses and soft-pulse sequences in a scalable fashion, without
the need for solving the full Hamiltonian.  Soft (narrow-band) pulses
are indispensable for their selectivity and reduced coupling to
environmental modes, which in turn suppresses signal distortions and
heating.  Use of high order pulses is especially efficient if one
interaction (e.g., the Ising term) is dominant.  High order pulse
sequences generally offer better accuracy and can dramatically reduce
the decoherence due to coupling with low-frequency environment.

{\bf Acknowledgments}.  The authors are grateful to Mark Dykman and
Daniel Lidar for encouragement and illuminating discussions.



\begin{thebibliography}{31}
\expandafter\ifx\csname natexlab\endcsname\relax\def\natexlab#1{#1}\fi
\expandafter\ifx\csname bibnamefont\endcsname\relax
  \def\bibnamefont#1{#1}\fi
\expandafter\ifx\csname bibfnamefont\endcsname\relax
  \def\bibfnamefont#1{#1}\fi
\expandafter\ifx\csname citenamefont\endcsname\relax
  \def\citenamefont#1{#1}\fi
\expandafter\ifx\csname url\endcsname\relax
  \def\url#1{\texttt{#1}}\fi
\expandafter\ifx\csname urlprefix\endcsname\relax\def\urlprefix{URL }\fi
\providecommand{\bibinfo}[2]{#2}
\providecommand{\eprint}[2][]{\url{#2}}

\bibitem[{\citenamefont{Slichter}(1992)}]{slichter-book}
\bibinfo{author}{\bibfnamefont{C.~P.} \bibnamefont{Slichter}},
  \emph{\bibinfo{title}{Principles of Magnetic Resonance}}
  (\bibinfo{publisher}{Springer-Verlag}, \bibinfo{address}{New York},
  \bibinfo{year}{1992}), \bibinfo{edition}{3rd} ed.; 
\bibinfo{author}{\bibfnamefont{R.}~\bibnamefont{Freeman}},
  \bibinfo{journal}{Progr. NMR Spectr.} \textbf{\bibinfo{volume}{32}},
  \bibinfo{pages}{59} (\bibinfo{year}{1998});
%
\bibinfo{author}{\bibfnamefont{L.~M.~K.} \bibnamefont{Vandersypen}}
  \bibnamefont{and} \bibinfo{author}{\bibfnamefont{I.~L.}
  \bibnamefont{Chuang}},
\bibinfo{journal}{Rev. Mod. Phys.}
  \textbf{\bibinfo{volume}{76}}, \bibinfo{pages}{1037} (\bibinfo{year}{2004}).

\bibitem[{\citenamefont{Vandersypen et~al.}(2001)\citenamefont{Vandersypen,
  Steffen, Breyta, Yannoni, Sherwood, and Chuang}}]{chuang-2001}
\bibinfo{author}{\bibfnamefont{L.~M.~K.} \bibnamefont{Vandersypen}},
  \bibinfo{author}{\bibfnamefont{M.}~\bibnamefont{Steffen}},
  \bibinfo{author}{\bibfnamefont{G.}~\bibnamefont{Breyta}},
  \bibinfo{author}{\bibfnamefont{C.~S.} \bibnamefont{Yannoni}},
  \bibinfo{author}{\bibfnamefont{M.~H.} \bibnamefont{Sherwood}},
  \bibnamefont{and} \bibinfo{author}{\bibfnamefont{I.~L.}
  \bibnamefont{Chuang}}, \bibinfo{journal}{Nature}
  \textbf{\bibinfo{volume}{414}}, \bibinfo{pages}{883} (\bibinfo{year}{2001});
%
\bibinfo{author}{\bibfnamefont{C.~S.} \bibnamefont{Yannoni}},
  \bibinfo{author}{\bibfnamefont{M.~H.} \bibnamefont{Sherwood}},
  \bibinfo{author}{\bibfnamefont{D.~C.} \bibnamefont{Miller}},
  \bibinfo{author}{\bibfnamefont{I.~L.} \bibnamefont{Chuang}},
  \bibinfo{author}{\bibfnamefont{L.~M.~K.} \bibnamefont{Vandersypen}},
  \bibnamefont{and} \bibinfo{author}{\bibfnamefont{M.~G.}
  \bibnamefont{Kubinec}}, \bibinfo{journal}{App. Phys. Lett.}
  \textbf{\bibinfo{volume}{75}}, \bibinfo{pages}{3563} (\bibinfo{year}{1999});
%
\bibinfo{author}{\bibfnamefont{G.~M.} \bibnamefont{Leskowitz}},
  \bibinfo{author}{\bibfnamefont{N.}~\bibnamefont{Ghaderi}},
  \bibinfo{author}{\bibfnamefont{R.~A.} \bibnamefont{Olsen}}, \bibnamefont{and}
  \bibinfo{author}{\bibfnamefont{L.~J.} \bibnamefont{Muller}},
  \bibinfo{journal}{J. Chem. Phys.} \textbf{\bibinfo{volume}{119}},
  \bibinfo{pages}{1643} (\bibinfo{year}{2003}).

\bibitem[{\citenamefont{Platzman et~al.}(2003)\citenamefont{Platzman, Dykman,
  and Seddighrad}}]{platzman-dykman-seddighrad}
\bibinfo{author}{\bibfnamefont{M.~I.} \bibnamefont{Dykman}},
\bibinfo{author}{\bibfnamefont{P.~M.} \bibnamefont{Platzman}}, 
  \bibnamefont{and}
  \bibinfo{author}{\bibfnamefont{P.}~\bibnamefont{Seddighrad}},
  \bibinfo{journal}{Phys. Rev. B} \textbf{\bibinfo{volume}{67}},
  \bibinfo{pages}{155402} (\bibinfo{year}{2003});
%
\bibinfo{author}{\bibfnamefont{A.~J.} \bibnamefont{Berkley}},
  \bibinfo{author}{\bibfnamefont{H.}~\bibnamefont{Xu}},
  \bibinfo{author}{\bibfnamefont{R.~C.} \bibnamefont{Ramos}},
  \bibinfo{author}{\bibfnamefont{M.~A.} \bibnamefont{Gubrud}},
  \bibinfo{author}{\bibfnamefont{F.~W.} \bibnamefont{Strauch}},
  \bibinfo{author}{\bibfnamefont{P.~R.} \bibnamefont{Johnson}},
  \bibinfo{author}{\bibfnamefont{J.~R.} \bibnamefont{Anderson}},
  \bibinfo{author}{\bibfnamefont{A.~J.} \bibnamefont{Dragt}},
  \bibinfo{author}{\bibfnamefont{C.~J.} \bibnamefont{Lobb}}, \bibnamefont{and}
  \bibinfo{author}{\bibfnamefont{F.~C.} \bibnamefont{Wellstood}},
  \bibinfo{journal}{Science} \textbf{\bibinfo{volume}{300}},
  \bibinfo{pages}{1548} (\bibinfo{year}{2003}).

\bibitem[{\citenamefont{Weiman}(1999)}]{weiman-1999}
\bibinfo{author}{\bibfnamefont{C.~E.} \bibnamefont{Weiman}},
  \bibinfo{journal}{Rev. Mod. Phys.} \textbf{\bibinfo{volume}{71}},
  \bibinfo{pages}{S253} (\bibinfo{year}{1999}).

\bibitem[{\citenamefont{Osborne and Coontz}(2002)}]{osborne-coontz-2002}
\bibinfo{author}{\bibfnamefont{I.}~\bibnamefont{Osborne}} \bibnamefont{and}
  \bibinfo{author}{\bibfnamefont{R.}~\bibnamefont{Coontz}},
  \bibinfo{journal}{Science} \textbf{\bibinfo{volume}{298}},
  \bibinfo{pages}{1353} (\bibinfo{year}{2002}).

\bibitem[{\citenamefont{Goswami}(2003)}]{goswami-review}
\bibinfo{author}{\bibfnamefont{D.}~\bibnamefont{Goswami}},
  \bibinfo{journal}{Phys. Rep.} \textbf{\bibinfo{volume}{374}},
  \bibinfo{pages}{385} (\bibinfo{year}{2003}).

\bibitem[{\citenamefont{Neogi et~al.}(1999)\citenamefont{Neogi, Yoshida,
  Mozume, and Wada}}]{laser-data-comm}
\bibinfo{author}{\bibfnamefont{A.}~\bibnamefont{Neogi}},
  \bibinfo{author}{\bibfnamefont{H.}~\bibnamefont{Yoshida}},
  \bibinfo{author}{\bibfnamefont{T.}~\bibnamefont{Mozume}}, \bibnamefont{and}
  \bibinfo{author}{\bibfnamefont{O.}~\bibnamefont{Wada}},
  \bibinfo{journal}{Opt. Commun.} \textbf{\bibinfo{volume}{159}},
  \bibinfo{pages}{225} (\bibinfo{year}{1999}).

\bibitem[{\citenamefont{Mehta et~al.}(1999)\citenamefont{Mehta, Rief, Spudich,
  Smith, and Simmons}}]{bio-laser}
\bibinfo{author}{\bibfnamefont{A.~D.} \bibnamefont{Mehta}},
  \bibinfo{author}{\bibfnamefont{M.}~\bibnamefont{Rief}},
  \bibinfo{author}{\bibfnamefont{J.~A.} \bibnamefont{Spudich}},
  \bibinfo{author}{\bibfnamefont{D.~A.} \bibnamefont{Smith}}, \bibnamefont{and}
  \bibinfo{author}{\bibfnamefont{R.~M.} \bibnamefont{Simmons}},
  \bibinfo{journal}{Science} \textbf{\bibinfo{volume}{283}},
  \bibinfo{pages}{1689} (\bibinfo{year}{1999}).

\bibitem[{\citenamefont{Abramovich and Vega}(1993)}]{abramovich-vega-1993}
\bibinfo{author}{\bibfnamefont{D.}~\bibnamefont{Abramovich}} \bibnamefont{and}
  \bibinfo{author}{\bibfnamefont{S.}~\bibnamefont{Vega}}, \bibinfo{journal}{J.
  Magn. Res. A} \textbf{\bibinfo{volume}{105}}, \bibinfo{pages}{30}
  (\bibinfo{year}{1993}).

\bibitem[{\citenamefont{Warren}(1981)}]{warren-herm}
\bibinfo{author}{\bibfnamefont{W.~S.} \bibnamefont{Warren}},
  \bibinfo{journal}{J. Chem. Phys.} \textbf{\bibinfo{volume}{81}},
  \bibinfo{pages}{5437} (\bibinfo{year}{1981});
%
\bibinfo{author}{\bibfnamefont{C.}~\bibnamefont{Bauer}},
  \bibinfo{author}{\bibfnamefont{R.}~\bibnamefont{Freeman}},
  \bibinfo{author}{\bibfnamefont{T.}~\bibnamefont{Frenkiel}},
  \bibinfo{author}{\bibfnamefont{J.}~\bibnamefont{Keeler}}, \bibnamefont{and}
  \bibinfo{author}{\bibfnamefont{A.~J.} \bibnamefont{Shaka}},
  \bibinfo{journal}{J. Mag. Res.} \textbf{\bibinfo{volume}{58}},
  \bibinfo{pages}{442} (\bibinfo{year}{1984});
%
\bibinfo{author}{\bibfnamefont{H.}~\bibnamefont{Geen}} \bibnamefont{and}
  \bibinfo{author}{\bibfnamefont{R.}~\bibnamefont{Freeman}},
  \bibinfo{journal}{J. Mag. Res.} \textbf{\bibinfo{volume}{93}},
  \bibinfo{pages}{93} (\bibinfo{year}{1991}).

\bibitem[{\citenamefont{Sanders et~al.}(1999)\citenamefont{Sanders, Kim, and
  Holton}}]{sanders-1999-complex-instruct}
\bibinfo{author}{\bibfnamefont{G.~D.} \bibnamefont{Sanders}},
  \bibinfo{author}{\bibfnamefont{K.~W.} \bibnamefont{Kim}}, \bibnamefont{and}
  \bibinfo{author}{\bibfnamefont{W.~C.} \bibnamefont{Holton}},
  \bibinfo{journal}{Phys. Rev. A} \textbf{\bibinfo{volume}{59}},
  \bibinfo{pages}{1098} (\bibinfo{year}{1999});
%
\bibinfo{author}{\bibfnamefont{A.~O.} \bibnamefont{Niskanen}},
  \bibinfo{author}{\bibfnamefont{J.~J.} \bibnamefont{Vartiainen}},
  \bibnamefont{and} \bibinfo{author}{\bibfnamefont{M.~M.}
  \bibnamefont{Salomaa}}, \bibinfo{journal}{Phys. Rev. Lett.}
  \textbf{\bibinfo{volume}{90}}, \bibinfo{pages}{197901}
  (\bibinfo{year}{2003}).

\bibitem[{\citenamefont{Steffen et~al.}(2003)\citenamefont{Steffen, Martinis,
  and Chuang}}]{steffen-2003}
\bibinfo{author}{\bibfnamefont{M.}~\bibnamefont{Steffen}},
  \bibinfo{author}{\bibfnamefont{J.~M.} \bibnamefont{Martinis}},
  \bibnamefont{and} \bibinfo{author}{\bibfnamefont{I.~L.}
  \bibnamefont{Chuang}}, \bibinfo{journal}{Phys. Rev. B}
  \textbf{\bibinfo{volume}{68}}, \bibinfo{pages}{224518} (\bibinfo{year}{2003}).

\bibitem[{\citenamefont{Domb and Green}(1974)}]{Domb-Green-book}
\bibinfo{editor}{\bibfnamefont{C.}~\bibnamefont{Domb}} \bibnamefont{and}
  \bibinfo{editor}{\bibfnamefont{M.~S.} \bibnamefont{Green}}, eds.,
  \emph{\bibinfo{title}{Phase transitions and critical phenomena}},
  vol.~\bibinfo{volume}{3} (\bibinfo{publisher}{Academic},
  \bibinfo{address}{London}, \bibinfo{year}{1974}); 
%
\bibinfo{author}{\bibfnamefont{J.}~\bibnamefont{G.~A.~Baker}},
  \emph{\bibinfo{title}{Quantitative theory of critical phenomena}}
  (\bibinfo{publisher}{Academic}, \bibinfo{address}{San Diego},
  \bibinfo{year}{1990}).

\bibitem{endnote} 
We also designed several
  narrow-band shapes replacing the entire refocusing sequences of
  various order with respect to Ising coupling $J_z$.  The advantage
  of individual pulses is the flexibility in cancelling the additional
  interactions, as well as the ability of using known high-order
  sequences\cite{brown-harrow-chuang-2004}.

\bibitem[{\citenamefont{Galassi et~al.}(2003)\citenamefont{Galassi, Davies,
  Theiler, Gough, Jungman, Booth, and Rossiet}}]{gsl-ref}
\bibinfo{author}{\bibfnamefont{M.}~\bibnamefont{Galassi}},
  \bibinfo{author}{\bibfnamefont{J.}~\bibnamefont{Davies}},
  \bibinfo{author}{\bibfnamefont{J.}~\bibnamefont{Theiler}},
  \bibinfo{author}{\bibfnamefont{B.}~\bibnamefont{Gough}},
  \bibinfo{author}{\bibfnamefont{G.}~\bibnamefont{Jungman}},
  \bibinfo{author}{\bibfnamefont{M.}~\bibnamefont{Booth}}, \bibnamefont{and}
  \bibinfo{author}{\bibfnamefont{F.}~\bibnamefont{Rossiet}},
  \emph{\bibinfo{title}{{GNU} {S}cientific {L}ibrary Reference Manual}},
  \bibinfo{edition}{2nd} ed. (\bibinfo{year}{2003}), \bibinfo{note}{{N}etwork
  Theory Ltd.}, \url{http://www.gnu.org/software/gsl/}.

\bibitem[{\citenamefont{Brown et~al.}(2004)\citenamefont{Brown, Harrow, and
  Chuang}}]{brown-harrow-chuang-2004}
\bibinfo{author}{\bibfnamefont{K.~R.} \bibnamefont{Brown}},
  \bibinfo{author}{\bibfnamefont{A.~W.} \bibnamefont{Harrow}},
  \bibnamefont{and} \bibinfo{author}{\bibfnamefont{I.~L.}
  \bibnamefont{Chuang}}, \bibinfo{journal}{Phys. Rev. A}
\textbf{\bibinfo{volume}{70}},  \bibinfo{pages}{052318}
  (\bibinfo{year}{2004});
%
\bibinfo{author}{\bibfnamefont{K.}~\bibnamefont{Khodjasteh}} \bibnamefont{and}
  \bibinfo{author}{\bibfnamefont{D.~A.} \bibnamefont{Lidar}}
   \bibinfo{note}{unpublished, quant-ph/0408128} (\bibinfo{year}{2004}).

\bibitem[{\citenamefont{Wimperis}(1994)}]{wimperis-1994}
\bibinfo{author}{\bibfnamefont{S.}~\bibnamefont{Wimperis}},
  \bibinfo{journal}{J. Magn. Res., Ser. A} \textbf{\bibinfo{volume}{109}},
  \bibinfo{pages}{221} (\bibinfo{year}{1994}).

\bibitem[{\citenamefont{Dykman}(1979)}]{dykman-1979}
\bibinfo{author}{\bibfnamefont{M.~I.}~\bibnamefont{Dykman}}, 
\bibinfo{journal}{Fiz. Nizk. Temp.} \textbf{\bibinfo{volume}{5}},
  \bibinfo{pages}{186} (\bibinfo{year}{1979})
  [\bibinfo{journal}{Sov. J. Low Temp. Phys.} \textbf{\bibinfo{volume}{5}},
  \bibinfo{pages}{89} (\bibinfo{year}{1979})].

\bibitem[{\citenamefont{Kofman and Kurizki}(2001)}]{kofman-kurizki-2001}
\bibinfo{author}{\bibfnamefont{A.~G.} \bibnamefont{Kofman}} \bibnamefont{and}
  \bibinfo{author}{\bibfnamefont{G.}~\bibnamefont{Kurizki}},
  \bibinfo{journal}{Phys. Rev. Lett.} \textbf{\bibinfo{volume}{87}},
  \bibinfo{pages}{270405} (\bibinfo{year}{2001});
%
  {\em ibid.} \textbf{\bibinfo{volume}{93}},
  \bibinfo{pages}{130406} (\bibinfo{year}{2004}).

\bibitem[{\citenamefont{Shiokawa and Lidar}(2004)}]{shiokawa-lidar-2004}
\bibinfo{author}{\bibfnamefont{K.}~\bibnamefont{Shiokawa}} \bibnamefont{and}
  \bibinfo{author}{\bibfnamefont{D.~A.} \bibnamefont{Lidar}},
  \bibinfo{journal}{Phys. Rev. A} \textbf{\bibinfo{volume}{69}},
  \bibinfo{pages}{030302(R)} (\bibinfo{year}{2004});
%
\bibinfo{author}{\bibfnamefont{L.}~\bibnamefont{Faoro}} \bibnamefont{and}
  \bibinfo{author}{\bibfnamefont{L.}~\bibnamefont{Viola}},
  \bibinfo{journal}{Phys. Rev. Lett.} \textbf{\bibinfo{volume}{92}},
  \bibinfo{pages}{117905} (\bibinfo{year}{2004}).
%

\bibitem[{\citenamefont{Pryadko}(2005)}]{pryadko-qke}
  \bibinfo{author}{\bibfnamefont{L.~P.} \bibnamefont{Pryadko}}
  \bibnamefont{et al.},
   \bibinfo{note}{unpublished} (\bibinfo{year}{2005}).

\bibitem{facchi-nakazato-2004}
  \bibinfo{author}{\bibfnamefont{P.}~\bibnamefont{Facchi}},
  \bibinfo{author}{\bibfnamefont{S.}~\bibnamefont{Tasaki}},
  \bibinfo{author}{\bibfnamefont{S.}~\bibnamefont{Pascazio}},
  \bibinfo{author}{\bibfnamefont{H.}~\bibnamefont{Nakazato}}, 
  \bibinfo{author}{\bibfnamefont{A.}~\bibnamefont{Tokuse}}, \bibnamefont{and}
  \bibinfo{author}{\bibfnamefont{D.~A.}~\bibnamefont{Lidar}},
  \bibinfo{note}{unpublished, quant-ph/0403205}
  (\bibinfo{year}{2004}).
\end{thebibliography}
\end{document}